\documentclass[]{article}
\usepackage{txfonts}
\usepackage[T2A]{fontenc}
\usepackage[utf8]{inputenc}
\usepackage[english,russian]{babel}
\usepackage[a4paper]{geometry}
\usepackage[hyper]{amsbib}
\usepackage{cite}

\begin{document}

\selectlanguage{english}

\twocolumn[{%

\small{Physics -- Uspekhi 67 (4) 405--416 (2024); doi: 10.3367/UFNe.2023.04.039575}\\
{\it CONFERENCES AND SYMPOSIA} \\[0.5em]
\LARGE {\bf Large-format imaging systems based\\on solid-state detectors in optical astronomy}

\small{V V Vlasyuk$^{1,a}$, I V Afanasieva$^{1,b}$, V I Ardilanov$^{1,c}$, V A Murzin$^{1,d}$, \\N G Ivaschenko$^{1,e}$, M A Pritychenko$^{1,f}$, S N Dodonov$^{1,g}$} \\[0.5em]
\footnotesize{ $^{(1)}$Special Astrophysical Observatory, Russian Academy of Sciences, 369167 Nizhniy Arkhyz, \\Zelenchukskiy region, Karachai-Cherkessian Republic, Russian Federation\\
E-mail:$^{(a)}$vvlassao@gmail.com, $^{(b)}$riv615@gmail.com, $^{(c)}$valery@sao.ru, $^{(d)}$vamur@sao.ru, \\$^{(e)}$ivanick@sao.ru, $^{(f)}$pma@sao.ru, $^{(g)}$dodo@sao.ru}\\[0.1em]

 \normalsize
}]

\noindent{\small \bf \underline{Abstract.} The development of technologies for creating various types of solid-state detectors for optical astronomy is reviewed. The principles of designing astronomical photodetecting systems with large-format sensors based on charge-coupled device (CCD) and complementary metal oxide semiconductor (CMOS) structures are analyzed. Examples of the most advanced projects to which they have been applied are given. The history of the creation of optical detectors for telescopes operated in Russia is described, and a brief description and characteristics of the developed systems are provided. The results of testing in real research are displayed. The prospects for creating large-format systems based on CCD and CMOS detectors manufactured in Russia and abroad are discussed.}
\vspace{.1\baselineskip}

\vspace{.1\baselineskip}

\vspace{1.1\baselineskip}

%=================================================================
\noindent{\large \textbf{1. Introduction}}\vspace{.7\baselineskip}

\noindent{This review is not intended to describe the entire variety of imaging systems based on solid-state detectors that have been used in astronomical research during the more than 40 years since the first instruments were introduced. We outline the basic concepts and technical solutions that allowed once modest, small-format, and not user-friendly devices to become almost ideal light sensors with virtually unlimited capabilities. Our review covers both CCD (Charge Coupled Device) and CMOS (Complementary Metal-Oxide­Semiconductor) detectors.}

Until the last decade, CCD detectors had been undisputed leaders in all astronomical projects involving the use of large­sized light detectors that did not require high temporal resolution.

Currently, a boom is being witnessed in the development of large-format CMOS detectors, which are replacing CCD detectors in a number of applications, both as individual monolithic devices and as mosaic systems. And yet, for a number of astrophysical problems, CCD detectors are still the main `workhorse.'

The developments carried out at the Special Astrophysical Observatory of the Russian Academy of Sciences (SAO RAS) since the mid-1980s have generally followed the main global trends, adjusted for the availability of certain technologies and the financial aspects of the issue kept in mind.

%=================================================================
\vspace{.9\baselineskip}
\noindent{\large \textbf{2. Charge-coupled devices ---\\from the first samples to giant `ideal' products}}\vspace{.7\baselineskip}

\noindent{\textbf{2.1 Front-illuminated CCD}}

\noindent{The principle of CCD operation based on the charge transfer from one semiconductor cell to another was discovered by W~Boyle and D~Smith~\cite{Boyle1970} working at Bell Laboratories (USA), and initially did not imply promising astronomical development: the authors studied problems of communication and signal transmission. The first astronomical application occurred in 1974, when a still imperfect 100~$\times$~100 element device manufactured by Fairchild Semiconductor was used to obtain an image of the Moon by an eight-inch telescope.}

Just two years later, such a device was professionally used for the first time to obtain images of Jupiter, Saturn, and Uranus on the 61-inch telescope at Mount Bigelow Observatory in Arizona~\cite{Smith1976}.

\begin{figure}
\centering
\includegraphics[width=.87\linewidth]{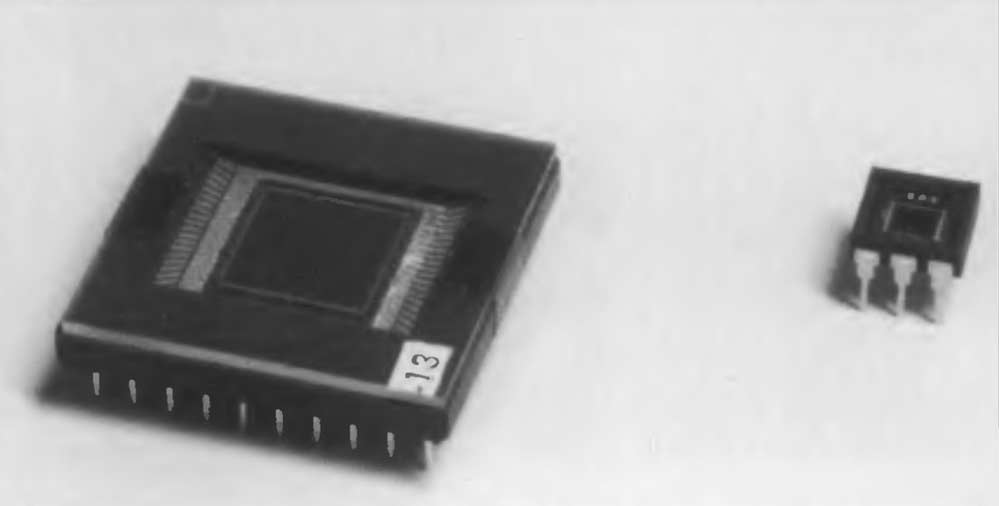}
\caption{\small {First commercially available CCDs: TH7895 (Thomson) consisting of 262,164 elements (left) and TI211 (Texas Instruments) consisting of 31,680 elements (right).}}
\label{fig_01}
\end{figure}

Since 1976, the five-meter Hale telescope of the Palomar Observatory and the telescopes of the Kitt Peak National Observatory and the Sierra Tololo Inter-American Observatory (since 1978) have been regularly using Texas Instruments and RCA Corporation detectors with a format of 400~$\times$~400 and 320~$\times$~580 elements, cooled with liquid nitrogen to temperatures below –120°C~\cite{Young1978, Pritchet1979, Oke1981, Goad1981, Perryman1982}. Nevertheless, the readout noise of such devices exceeds 40--50~e$^-$. A general view of the first-generation devices is displayed in Fig.~\ref{fig_01}.

Although the manufacturing technology of the devices was further developed and the accuracy and noise parameters improved, the dimensions of the detectors remained relatively mo-dest almost until the late 1980s~\cite{Janesick1987}, when competition between manufacturers began for growing orders from consumers, astronomers, and developers of various space missions. By 1990, the detector format of 1024~$\times$~1024 elements had become commonplace, and three technology leaders, Kodak, Tektronix, and Ford Aerospace, began producing detectors of an even larger format consisting of 2048~$\times$~2048 elements~\cite{Bredthauer1989}. At this stage, the Ford Aerospace Corporation took the lead, having mastered, in collaboration with Photometrics, the production of devices with a format of 4096~$\times$~4096 7.5-$\mu$m elements. They managed to achieve a record low readout noise --- less than 1~e$^-$ using repeated nondestructive readout of the charge from the detector element. Practical noise values were about 2--3~e$^-$ at a detector readout rate of about 50~kHz (with an increase in rate by an order of magnitude, noise usually increases by a factor of one and a half to two)~\cite{Janesick1990}.

Progress in the creation of large-format monolithic detectors in the 1990s is associated with Semiconductor Technology Associates (STA) and Fairchild Imaging.

Actually, Fairchild Imaging originated from Fairchild Semiconductor, one of the pioneer companies that created the first CCDs as early as 1973. Subsequently going through a series of mergers and acquisitions, it inherited specialized divisions from Loral and Ford Aerospace, and by 2000 concentrated on the production of large monolithic detectors for industrial, medical, and scientific applications~\cite{Vu2004}.

\begin{figure}
\centering
\includegraphics[width=.87\linewidth]{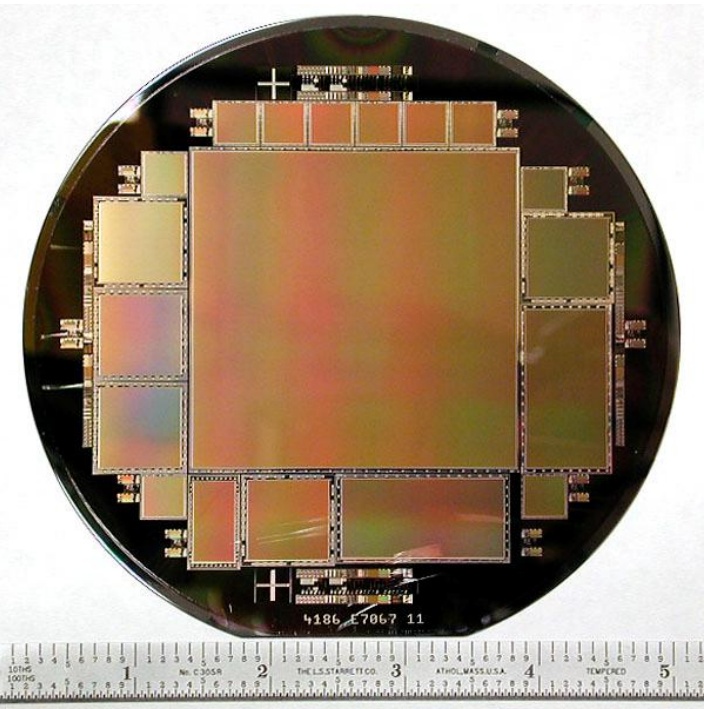}
\caption{\small{View of Fairchild CCD486 on a silicon wafer with a diameter of 125 mm.}}
\label{fig_02}
\end{figure}

As a result, a few years later, the company began mass production of the largest CCDs that can be accommodated on a silicon wafer with a diameter of 125 mm --- from CCD485/486 with a matrix or 4096~$\times$~4096 15-$\mu$m elements (shown in Fig.~\ref{fig_02}) to CCD595 with a matrix of 9216~$\times$~9216 8.75-$\mu$m elements.

The entire developed line of devices, with the exception of the CCD486 (this model had the option of back illumination of the detector), like earlier developments, used front illumination technologies, in which part of the working surface is covered with electrodes and remains insensitive to incident radiation, which limits the quantum efficiency of registration to 45--50\%.

As of 2004, the CCD486 manufactured by Fairchild Imaging remained the world's largest commercially distributed CCD (devices like the CCD595 were produced in limited quantities and, most likely, for the US Government alone). It could simultaneously read a signal from all four corners with an excellent charge transfer efficiency of no worse than 0.999995 for each coordinate. Using special doping additives to silicon, the developers significantly reduced the thermal generation (dark) current to 0.02~e$^-$/s at a crystal temperature of $-60^\circ$C. The readout noise of the device was approximately 5~e$^-$ at a readout rate of 50 kHz and better than 10~e$^-$ at a rate of 1 MHz.

\begin{figure}
\centering
\includegraphics[width=.87\linewidth]{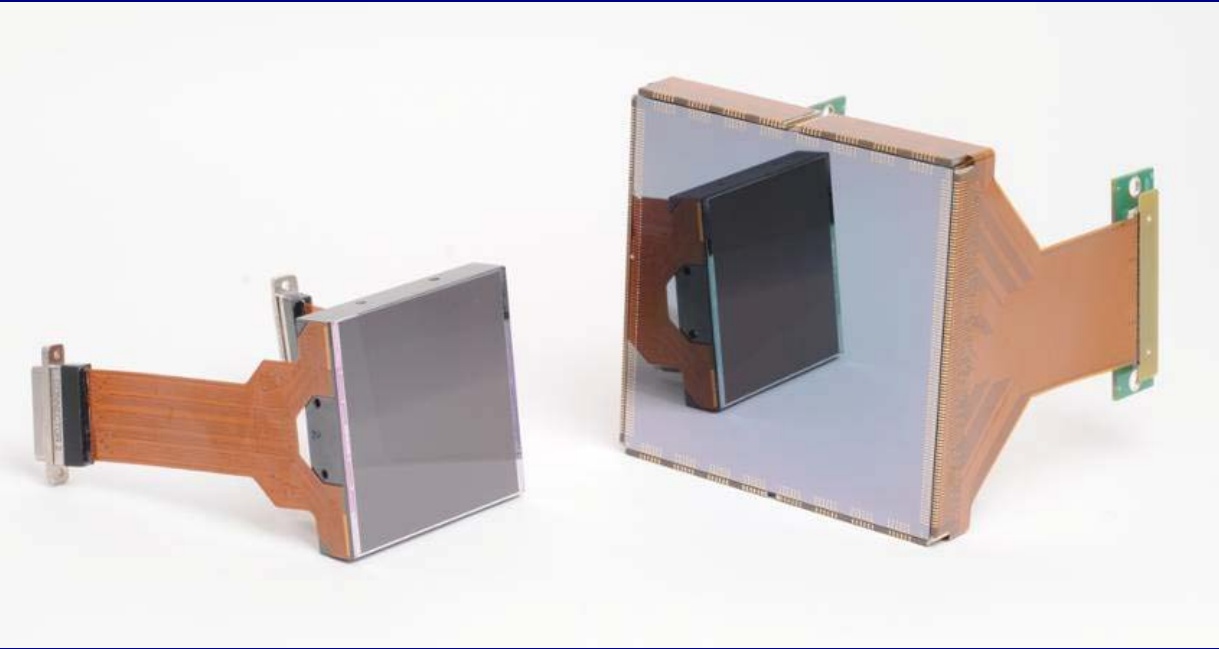}
\caption{\small{STA 1600B detector (right) compared with a 4K~$\times$~4K E2V CCD231 device.}}
\label{fig_03}
\end{figure}

Unlike Fairchild Imaging, a major company with a rich history, STA is a small company founded in 1999, which is the brainchild of a talented detector designer, R Bredtower. His experience in the development since 1975 of large-format detectors at Lockheed Martin, Loral, Ford Aerospace, and Rockwell for the Mars missions Pathfinder and Cassini and the Hubble Space Telescope enabled STA after 10 years of operations to produce the world's largest monolithic CCDs with excellent parameters. For example, the STA 1600B device with linear dimensions of 95$\times$95 mm, consisting of a matrix of 10560$\times$10560 9-$\mu$m elements~\cite{Bred2012} (shown in Fig.~\ref{fig_03} on the right), due to 16 low-noise outputs, provides a maximum readout rate of approximately 1 fps. Readout noise at low rates is approximately 3~e$^-$ and at a 1-MHz rate, the noise increases to 5--7~e$^-$.

Despite such excellent parameters, STA only managed to be invited to a few large astronomical projects~\cite{Dorland2005, Kwiat2008, Marin2012}, continuing to replicate its successful developments.

Among the companies operating in this technological field outside the United States, Sony Semiconductor (Japan) and Marconi, later EEV, then E2V Technologies (Great Britain), and now Teledyne e2v (USA) should be mentioned. While Sony made a significant contribution to improving CCD production technologies and then switched in the 2000s to the development of CMOS detectors, the UK-based company by the early 2010s had already won most of the market for large-format CCD detectors used in astronomy.

Returning to the beginning of the section, it should be noted that the accomplishments of the discoverers of CCDs were appreciated 40 years later: Boyle and Smith were awarded the Nobel Prize in Physics in 2009 for their contribution to the creation of charge-coupled device technology.

%=================================================================
\vspace{.7\baselineskip}
\noindent{\textbf{2.2 Introduction of back-illuminated CCDs}}

\noindent{The quantum response of photon detection in a CCD structure can be increased several fold (depending on the range, up to a factor of 3 to 5) provided that a thinner silicon layer is used and the light is directed `from the back side'. In this case, the loss of light due to absorption and reflection from the surface of the electrodes that form the electric field in the photosensitive layer is eliminated. In the wavelength range less than 450--500 nm, the detection efficiency will be many times better, whereas in the wavelength range exceeding 650--700 nm, the efficiency of the devices begins to be comparable due to the greater depth of light absorption. Since the surface of a pure silicon layer reflects almost 30\% of the incident light, the use of special multilayer anti-reflective coatings can significantly increase quantum efficiency.}

In back-illuminated CCDs, the image is focused on the back of the silicon substrate, bypassing the control electrode structure. However, to achieve (high) efficiency, the thick substrate on which the CCD is located should be almost completely thin-ned because only electrons generated near the control electrodes can be effectively collected. The thickness of the substrate of such CCD detectors does not exceed 10--12~$\mu$m.

This approach was first implemented in practice by M Les-ser from the University of Arizona in the second half of the 1990s \cite{Lesser1996, Lesser1998}. He used selective chemical etching, which ultimately yielded a thickness of about 10--20-$\mu$m with a low density of defects. Silicon thickness is controlled by the difference between the etching rates of the heavily doped substrate and the grown silicon layer. For application in large-format devices, this method was refined in collaboration with STA, Fairchild Imaging, and Imager Labs~\cite{Lesser2004}. The production of large-format devices apparently sets more stringent requirements for the pro-perties of the materials used, which inevitably increases the cost of manufacturing such products.

The joint work allowed Fairchild Imaging to implement this technology in the production of large-format instruments, in particular for CCD486~\cite{Lesser2001}, which became popular in a number of astronomical projects: KeplerCam~\cite{Szent2005}, LCOGT~\cite{Tufts2008}, and Lick AGN Monitoring Project~\cite{Bentz2009}.

Teledyne e2v has been even more successful in implementing the back-side illumination technique in its devices. The high quality of the instruments, well-established cooperation with camera manufacturers, and large production volumes allowed the company to become involved in most ongoing astronomical projects. We only mention a small part of the research in which single CCD231 devices manufactured by Teledyne e2v in the 4K~$\times$~4K format are used (shown on Fig.~\ref{fig_03} on the left). The list of these devices includes the PMAS integrated field spectrophotometer for the 1.5-m telescope of the Calar Alto Observatory~\cite{Roth2010}, the multimode 24-channel MUSE instrument for the VLT telescope~\cite{Reiss2012}, and the MEGARA field and multi-object spectrograph for the 10-m GTC telescope~\cite{Cast2012}.

%=================================================================
\vspace{.7\baselineskip}
\noindent{\textbf{2.3 Deep-depletion back-illuminated CCDs}}

\noindent{Attempts to create devices with back illumination and a sufficiently deep layer of depleted silicon to enhance the detection efficiency in the long-wavelength range began in the mid-1990s at the Lawrence Berkeley National Laboratory (LBNL) in California, USA~\cite{Holland1996}. Subsequent studies showed significant advantages of this type of imaging. A typical silicon thickness of 200--300~$\mu$m results in high quantum efficiency in the red part of the spectrum and makes it possible to significantly suppress interference effects in studies conducted in this range~\cite{Groom1999}. The response to an incident photon is clearly defined and depends on the time taken for the photogenerated holes to pass through the electric field, which extends across the entire thickness of the device~\cite{Holland1997, Groom99}.}

In 2004, Lawrence Berkeley National Laboratory, together with Canada-based Dalsa Semiconductor, announced the development of a technology for the production of large-format devices for astronomical tasks that feature the necessary properties: enhanced sensitivity in the red part of the spectrum and a virtually complete absence of interference patterns~\cite{Holland2003, Bebek2004}. Devices of the 2K~$\times$~4K format with a 15~$\mu$m element yielded readout noise or 4.5~e$^-$, dark current of about 7~e$^-$/hour per element, charge transfer efficiency better than 0.999995, and a potential well depth of more than 160~Ke$^-$ at an operating temperature of $-140^\circ$C. The quantum efficiency was $\sim$ 60\% at $\lambda$~=~400 nm, more than 90\% at $\lambda$~=~700--900 nm, and 60\% at $\lambda$~=~1000 nm. Devices of this class in the format of 800~$\times$~1980 and 2048~$\times$~2048 elements were used for astronomical observations with telescopes at the Kitt Peak Observatory~\cite{Bebek_2004}.

The successful implementation of the technology of deep depletion of thick silicon substrates has encouraged developers in other countries. Two years later, Japanese developers announced the creation of a fully buttable 2K~$\times$~4K Hamamatsu Photonics instrument to create mosaics for the 8.2-m Subaru telescope. With a silicon layer thickness of 200~$\mu$m, a charge transfer efficiency better than 0.999995 and a readout noise of less than 5~e$^-$ at a readout speed of 150 kHz were attained. The root mean square value of charge diffusion did not exceed 7.5~$\mu$m, the flatness of the detector was better than 15-20~$\mu$m, and the dark current was several ~e$^-$/hour per element at an operating temperature of $-100^\circ$C~\cite{Kamata2006}.

At about the same time, E2V Technologies announced the development of its own version of a large-format CCD based on high-resistance silicon~\cite{Jorden2006} with a depletion depth of about 100~$\mu$m. The substrate thickness was up to 600~$\mu$m, and the voltage supplied to the electrodes was approximately 70~V. The studies were carried out for devices in the format of 2K~$\times$~512 and 2K~$\times$~4K elements. The abovementioned work laid the foundation for the line of deep-depletion CCD manufactured by this company.

%=================================================================
\vspace{.7\baselineskip}
\noindent{\textbf{2.4 Electron-multiplying CCDs}}

\noindent{The principle of electronic internal charge multiplication directly on the CCD crystal (more exactly, in its output register) was first implemented by E2V Technologies in 2001~\cite{Janesick2001, Jerram2001}. By modifying the output register of a CCD and applying increased voltage to its electrodes (up to 30--40~V instead of the standard 10~V), it is possible to accelerate photoelectrons as they pass through each element and increase the probability (up to 2\%) of generating a second electron by avalanche multiplication. After passing through several hundred elements, the final gain can be several hundred or several thousand. At the register output, the readout noise generated by the output amplifier is divided by this total gain, resulting in an effective readout noise of a fraction of an electron.}

Already in its first years, E2V Technologies manufactured a line of such devices under the name L3CCD (Low-Level-Light CCD) with sizes from 128~$\times$~128 to 1024~$\times$~1024 16-$\mu$m elements. Such detectors can have both front and back illumination (with a thin silicon structure). Another manufacturer, Texas Instruments, also offered detectors with very similar electron multiplication technology under the Impactron brand. They have many devices available in formats up to 1024~$\times$~1024 8-$\mu$m elements~\cite{Mackay2004}.

Due to reduction in readout noise, electronic multiplying CCDs (EM~CCDs) are an important step in improving the performance of detecting systems and have significant potential for certain astronomical applications. EM CCDs feature all the advantages of conventional CCDs in terms of quantum efficiency, charge transfer efficiency, dark current, cosmetic quality, etc.

The only significant disadvantage of EM~CCD is the distortion of the Poisson statistics of light as a result of the superposition of the multiplication process statistics associated with the probabilistic nature of the amplification mechanism based on the impact ionization phenomenon. The dispersion effect resulting from the stochastic multiplication process leads to an increase in the readout noise by a factor of $\sqrt{2}$~\cite{Robins}.

Thus, the back-illuminated EM CCD --- a device that combines both electron multiplication and back-illuminated technology – can be expected to provide both single-photon detection sensitivity (by eliminating readout noise) and maximum efficiency of photon conversion. A back-illuminated EM CCD can theoretically improve the signal-to-noise ratio by a factor of two or more (depending on wavelength) compared with a front-illuminated EM CCD.

Andor, the first commercial supplier of digital CCD cameras that used this innovative technology, manufactured the Andor Technology iXon series of scientific cameras distributed under the EM CCD brand. These devices are now actively used in scientific research that set stringent requirements for the readout noise and require high data acquisition rates: high frame-rate imaging to improve angular resolution \cite{Mackay2004}, fast photometry~\cite{Smith2004}, and high-resolution spectroscopy \cite{Daigle2004, Daigle2006, Wena2006, Ives2008}.

\begin{figure}
\centering
\includegraphics[width=.87\linewidth]{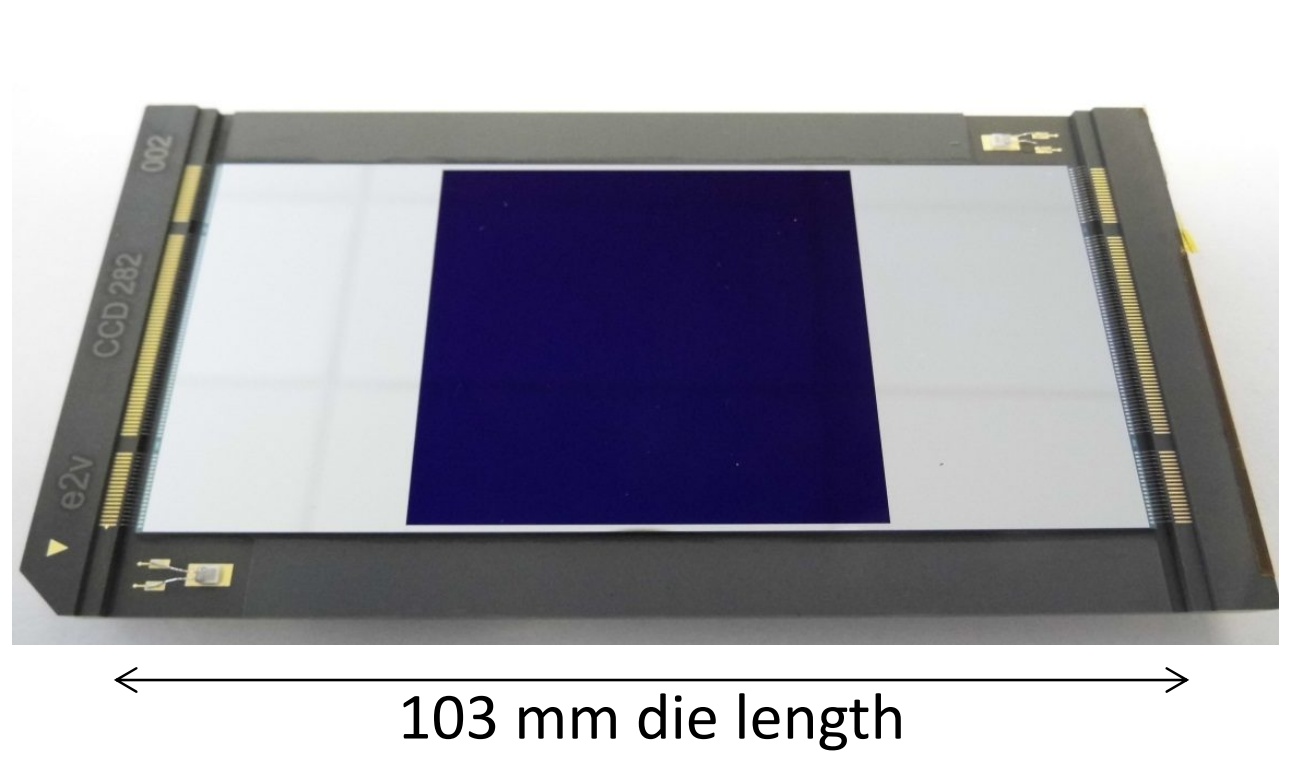}
\caption{\small{General view of the device with electron multiplication CCD282. Located to the left and right of the active area of the detector are shielded memory sections that provide frame-by-frame transfer of the accumulated image. The length of the entire crystal is 103 mm.}}
\label{fig_04}
\end{figure}

Subsequent developments in this area include the creation by E2V Technologies of a large-format detector with back illumination and electronic multiplication (BI EM CCD) CCD282 with the format of 4K~$\times$~4K 12-$\mu$m elements~\cite{Jorden2014}. An image of the device is displayed in Fig.~\ref{fig_04}.

This detector, which was commissioned by the University of Montreal, was used to create a scanning Fabry-Perot interferometer installed on the NTT telescope of the European Southern Observatory (ESO) and on larger telescopes~\cite{Gach2014, Daigle2018}. To increase the sampling rate of the detector, it uses a frame transfer mode and 8 parallel output nodes. Despite these dimensions, it provides a readout speed of 5~fps with sub-electron readout noise.

%=================================================================
\vspace{.7\baselineskip}
\noindent{\textbf{2.5 Buttable CCDs and close-butted mosaics in astronomy}}

\noindent{The maximum size of a single CCD detector is determined by the diameter of the silicon wafer that can be manufactured using existing technologies. In the 1990s, this value was 125 mm (see Fig.~\ref{fig_02}), while in the 2000s, it became possible to use high-quality plates with a diameter of 150 mm; in the last decade, their diameter has already reached 200 mm. To create imaging systems of a larger size, it will be necessary to create mosaic detectors, which led to the development of detectors with three- and four-way connectivity. Most currently produced models of CCD detectors have this feature (see, for example. Fig.~\ref{fig_03}).}

Among the large number of mosaic systems developed for astronomical tasks, we only emphasize the most significant developments that opened the way for subsequent projects.

One such system --- modest in size, but a breakthrough for its time --- is the wide-field camera of the Hubble Telescope, WF/PC (after modernization with the replacement of instruments in 1993, WFPC2), research on which began in 1990. Each modification used four devices of the 800~$\times$~800 element format (originally produced by Texas Instruments, later by Loral)~\cite{Holtzman1995}. The results obtained using this device enabled astronomers to take a new look at our Universe and are described in thousands of scientific publications. Due to unsolved problems of coupling of CCDs, the field was optically divided into individual devices.

\begin{figure}
\centering
\includegraphics[width=.9\linewidth]{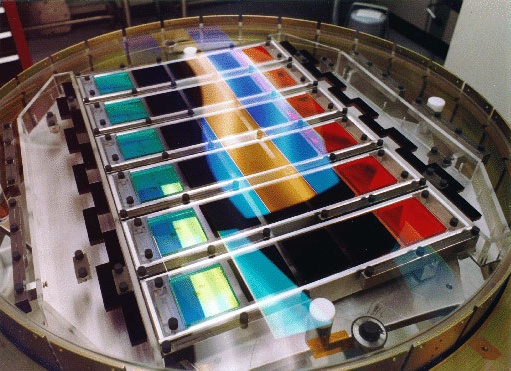}
\caption{\small{View of the Sloan Survey Telescope camera comprising 30~assembled CCD devices.}}
\label{fig_05}
\end{figure}

Preparations for a fully digital photometric and spectral survey of the sky were carried out in the 1990s, and in 2000, the Sloan Digital Sky Survey (SDSS) was launched on a specialized 2.5-m telescope located at the Apache Point Observatory (USA). The telescope was equipped with a wide-angle mosaic CCD camera (Fig.~\ref{fig_05}) and a pair of dual spectrographs with fiber optic input. A new feature of the telescope is a 3$^\circ$ focal plane (in linear measure, 0.65 m), which provides excellent image quality and low geometric distortion over a wide range of wavelengths (300--1060 nm).

The imaging camera collects photometric data using an array of 30 2048~$\times$~2048 SITe/Tektronix CCDs arranged in six columns of five CCDs each, aligned with the columns of the CCDs themselves. The detector element size is 24~$\mu$m, and the readout noise does not exceed 5~e$^-$ at the corresponding rate. During the process of shooting, the device operates in drift scanning mode: the camera slowly reads images, while images of transferred objects are moved along the columns of CCD detectors with the speed of rotation of the celestial sphere. Thus, the camera creates five images of a given object in all colors of the survey~\cite{Gunn1998}.

SDSS provided scientists with data for detailed studies of the distribution of luminous and non-luminous matter in the Universe: a photometrically and astrometrically calibrated digital image of $\pi$ steradians of the celestial sphere above 30$^\circ$ of galactic latitude in five optical bands and a spectroscopic survey of about $10^6$ of the brightest galaxies and $10^5$ of the brightest quasars discovered in the catalog of photometric objects~\cite{York2000,Gunn2006}.

Since the beginning of the 2000s, new technological capabilities, partially outlined earlier, initiated an almost explosive growth in the number of mosaic systems created for astronomical problems.

\begin{figure}
\centering
\includegraphics[width=.87\linewidth]{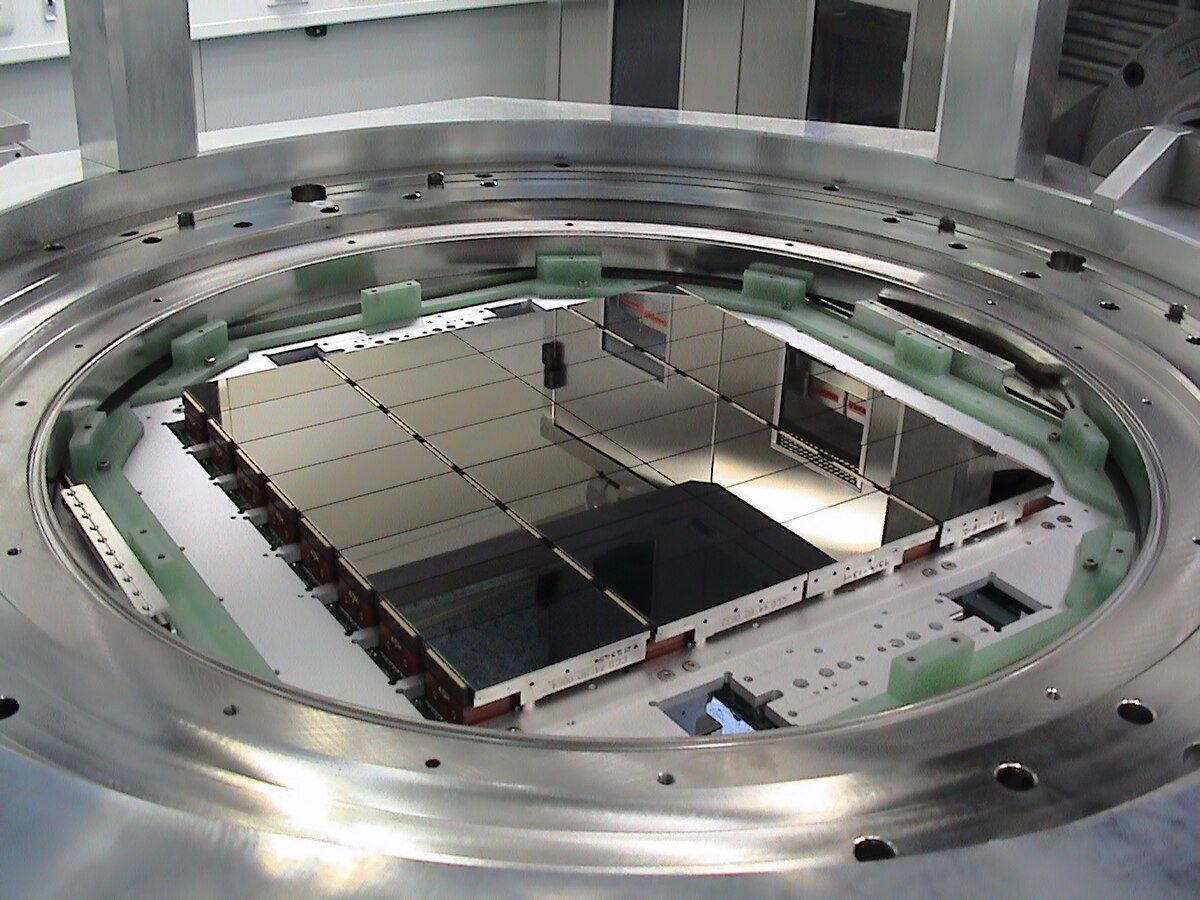}
\caption{\small{View of the OmegaCAM mosaic device.}}
\label{fig_06}
\end{figure}

For \,example, \,in \,2006, ESO \,announced \,the \,creation \,of OmegaCAM, \,an \,8~$\times$~4 \,mosaic of \,32 \,CCD44-80s \,produced by~e2v in the format of 2K~$\times$~4K 15-$\mu$m elements (Fig.~\ref{fig_06}). OmegaCAM was installed on the ESO VST 2.5-m survey telescope and provided a 1$^\circ$ field of view with an excellent quality of $0\mbox{.\kern -0.7ex\raisebox{.9ex}{\scriptsize$\prime\prime$}}21$ per element~\cite{Iwert2006}.

The Kepler space mission, which searched for and studied exoplanets around solar-type stars within an area of approximately 100~deg$^2$ using a 1-m aperture telescope, was launched into orbit in 2009~\cite{Caldwell2010, Koch2010}. For its time, it was equipped with the largest mosaic consisting of 42 CCD detectors produced by e2v in the format of 2200~$\times$~1024 27-$\mu$m elements, with a total size of 95 million elements. During the four years of the mission's operations, several thousand planets orbiting around other stars were discovered in its data.

At about the same time as the Kepler project, development of a camera to search for dark energy (DECam), intended for ground-based research at the primary focus of the 4-m telescope at the Sierra Tololo Observatory, began. It is an array of 62 high-resistance silicon detectors (BI DD) jointly produced by the Canadian Dalsa and LBNL in the format of 2K~$\times$~4K 15-$\mu$m elements~\cite{Cease2008}. The camera’s working field of 520 million elements, which has a linear size of approximately 0.5 m, allows simultaneous imaging of an area of 3~deg$^2$ on the celestial sphere.

As a result of the first DECam survey, conducted in 2013--2019, about 200 scientific papers were published, which made it possible to clarify the characteristics of the distribution of dark matter and compare them with the results obtained from the examination of cosmic microwave background radiation (in particular, \cite{DESC2016, Brout2019}). Research with this camera continues as part of other surveys (see, for example, \cite{Wang2017}).

One of the most important providers of up-to-date information about non-stationary objects in the Universe --- the first wide-angle telescope of the Pan-STARRS network --- started regular operations in 2010. It consists of a primary mirror with a diameter of 1.8 m and an aperture ratio of $f/4.4$ and a secondary mirror with a diameter of 0.9 m, providing a field of view exceeding 3$^\circ$. The optical design ensures low distortion and minimal vignetting, even at the edges of the working area. By now two telescopes of the originally planned network have been put into operation.

The Pan-STARRS-1 camera, which consists of $\sim$~1.4 billion elements, is a mosaic of 60 back-illuminated CCDs with a format of 4846~$\times$~4868 CCID58 elements, manufactured by the Lincoln Laboratory at the Massachusetts Institute of Technology~\cite{Tonry2008}. The size of the CCD detector elements is 10~$\mu$m, which corresponds to $0\mbox{.\kern -0.7ex\raisebox{.9ex}{\scriptsize$\prime\prime$}}258$ on the celestial sphere; the thickness of the silicon layer is 70~$\mu$m. All detectors are sampled using a specialized CCD controller; the readout time required to obtain a full frame is 7~s. The readout noise for all devices at this rate is approximately 5~e$^-$. Active, usable mosaic elements occupy approximately 80\% of the field of view~\cite{Magnier2020}. 

At the end of this section, two major projects should be mentioned, the implementation of which is expected in the near future.

JPCam (Javalambre Panoramic Camera) is a mosaic camera consisting of 1.2 billion elements. It is being created to be installed on the 2.6-m survey telescope of the Javalambre Observatory and includes 14 large-format CCD290-99 detectors manufactured by e2v, each containing 9.2K~$\times$~9.2K 10-$\mu$m elements. The cryostat for the mosaic was also developed by e2v. The camera to be installed at the Cassegrain focus of the telescope will provide a full field of view with a diameter of 3$^\circ$ and a scale of $0\mbox{.\kern -0.7ex\raisebox{.9ex}{\scriptsize$\prime\prime$}}23$ per element~\cite{Robbins2016}.

The main goal of this instrument is to conduct a photometric survey of the northern sky in 54 narrow-band filters to obtain, with unprecedented accuracy, the photometric redshifts of billions of faint galaxies over an area of approximately 8000 deg$^2$. Tests have shown that the entire mosaic is read within 10--15 s, and a readout noise of the order of 5~e$^-$ is realized~\cite{Marin2022}.

The camera of the Large Synoptic Survey Telescope (LSST) contains a mosaic of 3.2 billion elements. A decade-long endeavor to create the largest survey telescope, the LSST, is coming to an end. To implement the goal set for the project --- to provide a working field of 3$^\circ$ with a main mirror diameter of 8.4 m --- the developers had to use a three-mirror system with a very large aperture ratio of $f/1.234$ and a three-lens corrector~\cite{Angel2000}.

The optical design was optimized to provide a large field of view with extreme image quality over a wide wavelength range (320--1050 nm). The incident light is collected by an annular primary mirror with an outer diameter of 8.4 m and an inner diameter of 5 m, creating an effectively filled aperture $\sim$~6.4~m in size. The collected light is reflected onto a convex 3.4 m secondary mirror, then from it, onto a concave tertiary mirror with a diameter of 5 m, and finally onto three refractive camera lenses. The three-element refractive optic apparatus of the camera corrects chromatic aberrations caused by the need to use a thick cryostat entrance window and aligns the focal surface.

\begin{figure}
\centering
\includegraphics[width=0.9\linewidth]{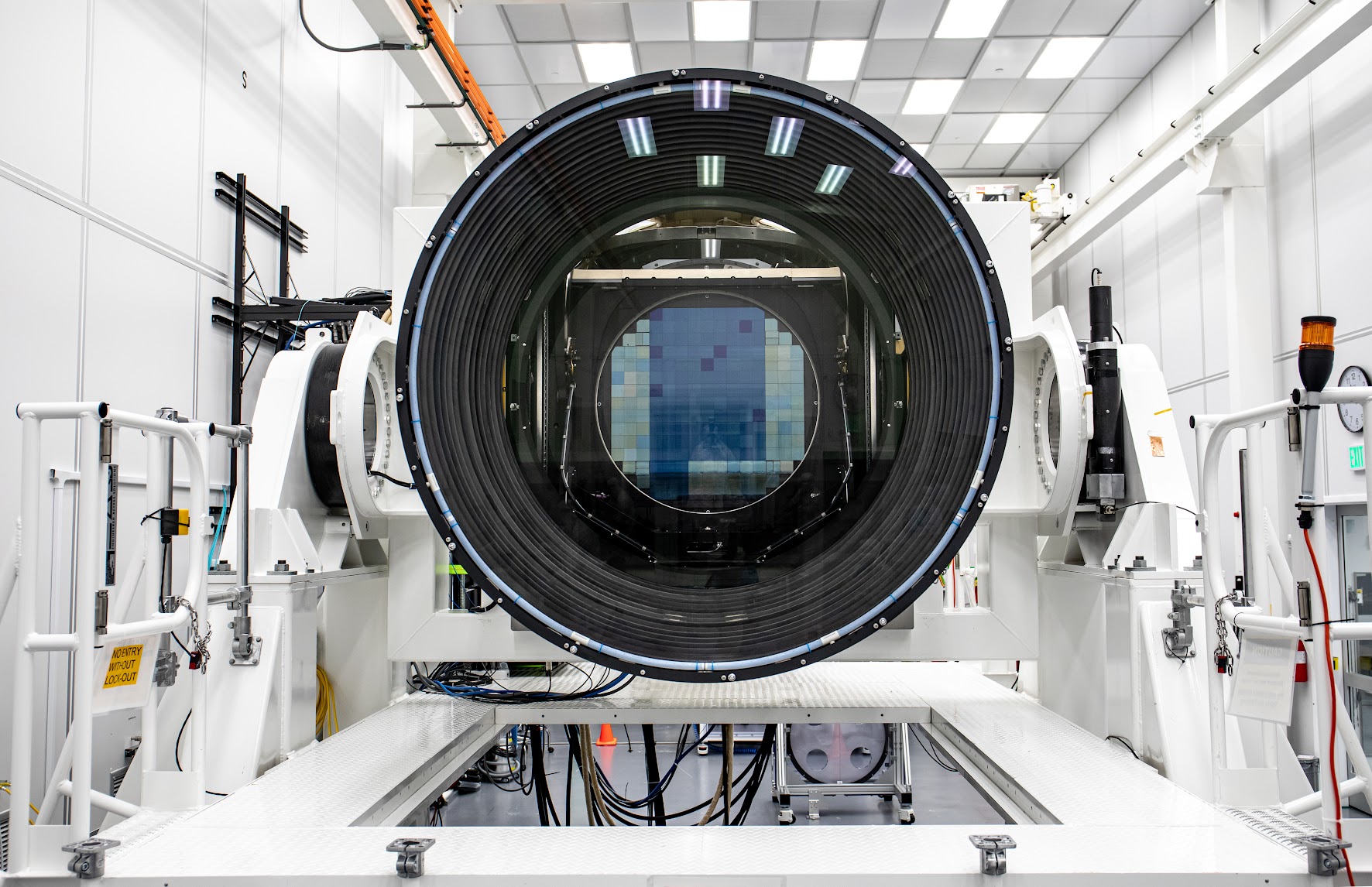}
\caption{\small{Mosaic camera for LSST consisting of 3.2 billion elements in an assembly shop.}}
\label{fig_07}
\end{figure}

The LSST camera contains a 3.2-billion-element mosaic consisting of 189 Teledyne e2v CCD250-82 CCDs in 4K~$\times$~4K format with 10-$\mu$m elements~\cite{Jorden2014} (shown in Fig.~\ref{fig_07}). This choice is determined by the need to span the entire field of view with a diameter of 0.64 m and a resolution of  $0\mbox{.\kern -0.7ex\raisebox{.9ex}{\scriptsize$\prime\prime$}}2$ per element. The CCD detectors themselves are made of high-resistance deep-depletion silicon with back illumination (BI DD CCD) and a highly segmented architecture, which enables the readout of the entire matrix in 2~s with a readout noise of at least 5~e$^-$. The detectors are grouped into 3~$\times$~3 blocks, each containing its own specialized electronics. These blocks are installed on a silicon carbide frame inside a vacuum cryostat with an individual thermal control system that maintains the operating temperature of the CCD devices at approximately 173 K (see~\cite{OConnor2008, Kahn2010, Doherty2014, Ivezic2019}). 

%=================================================================
\vspace{.9\baselineskip}
\noindent{\large \textbf{3. CMOS devices in astronomy}}\vspace{.7\baselineskip}

\noindent{\textbf{3.1 CMOS detector operation principle}}

\noindent{The idea of using a complementary metal oxide semiconductor in the technology for producing two-dimensional arrays of light-sensitive pixels is almost 40 years old. However, the first CMOS detectors were designed exclusively for use in hybrid infrared detectors and were therefore too exotic and expensive to gain widespread use. CMOS technology was excellent for the infrared region, where silicon is transparent and the elements are large. As a result, by the mid-1990s, large hybrid arrays with a format of up to 1K~$\times$~1K elements became available. Around the same time, CMOS technology reached sizes small enough to enable the creation of monolithic CMOS sensors with small elements and high performance. However, years had to pass before CMOS sensors became capable of reaching the high performance levels of CCDs. As of the mid-2000s, CMOS image sensors had occupied a significant part of the commercial sensor market and were present in a variety of consumer devices. In addition, hybrid CMOS technology was further developed in the form of multi-mega-element devices for both the infrared and visible ranges.}

Similar to CCDs, CMOS detectors use the photoelectric effect within the bulk of a semiconductor to convert photons into electrical charges. However, in contrast to CCDs, CMOS sensors can be used as a detector for materials other than silicon. In contrast, a number of various materials can be hybridized with a CMOS readout chip to provide sensitivity in the ultraviolet, visible, or infrared ranges.

Both technologies use a photodiode to generate and separate charges in each element, but then the differences emerge. During readout, the CCD chip moves the collected charge line by line to the output register, and then, in each row, all the charges move sequentially to a single output node, where one amplifier generates the corresponding output voltages. On the other hand, CMOS detectors contain an independent amplifier in each element that converts the stored charge into voltage, thus eliminating the need to transfer charge from one element to another. Instead, the voltages are multiplexed onto a common bus using on-chip CMOS switches. Analog or digital sensor outputs can be used, depending on whether a video output amplifier or an analog-to-digital converter is employed.

A significant advantage of CMOS sensor technology is its high flexibility. Small and simple elements with three or four transistors are used to detect the signal, whereas larger, more complex elements with hundreds of transistors provide analog-to-digital conversion and other image processing capabilities directly at the element level.

Typically, CMOS detectors do not require a mechanical shutter because the acquisition time is electronically controlled. An interesting feature is that the pixels can be read without destroying the signal of the built-in detector (nondestructive reading). This feature allows each element to be read more than once, thereby reducing readout noise by averaging multiple readings. Unlike a CCD, a CMOS sensor can be scanned in several different ways, including random access to any element at any time~\cite{Hoffman2005}.

CMOS detectors are divided into two classes: monolithic and hybrid. Monolithic CMOS sensors are image sensor chips that combine photodetectors and readout circuitry on the same silicon wafer. Compared with hybrid sensors, monolithic matrices are cheaper to manufacture, but their sensitivity is limited to the visible and near-infrared wavelength ranges. In a monolithic CMOS sensor, photodiodes share the elemental area with transistors. For this reason, the fill factor is always less than 100\%. In addition, front-illuminated CMOS detectors have limited sensitivity in the red range due to the relatively weak absorption of photons by silicon.

In a hybrid device optimized for the optical range, the p-i-n detector array consists of a thick ($\sim$~185~$\mu$m) active region of high-purity silicon located between p-type and n-type doped regions~\cite{Bai2000}. Detection usually occurs in the n-type doped region, whereas the side adjacent to the readout circuit is doped with p-type additives. A strong reverse bias is applied to the device, resulting in a strong electric field that separates electron-hole pairs created by photons absorbed in a high-resistivity region. The main part of the detector array is almost completely depleted, providing excellent quality at long wavelengths and high transmission characteristics.

%=================================================================
\vspace{.7\baselineskip}
\noindent{\textbf{3.2 First hybrid CMOS detectors for optical astronomy}}

\noindent{By the mid-2000s, an apparent technological leader in the production of CMOS detectors for a wide range of applications emerged: Rockwell Scientific Company (RSC), known to the astronomical community primarily for its line of HAWAII infrared detectors used in many infrared (IR) systems.}

For the optical range, RSC has developed two types of hybrid detectors: sensors made from HgCdTe with a removed substrate grown by molecular beam epitaxy and a p-i-n silicon matrix. While HgCdTe with a removed substrate provides continuous coverage of the visible and infrared ranges, a device based on a p-i-n matrix allows detection of only optical radiation, but at a lower cost and at nitrogen cooling temperatures.

Its quantum efficiency exceeds 80\% in the range of 500--900 nm and is more than 50\% in the range of 900--1000 nm due to the thickness of the silicon detector layer. Such devices feature a high percentage of operable elements ($>$ 99.99\%), readout noise of single devices of less than 13~e$^-$ at \,a \,clock \,frequency \,of \,100 kHz, \,and \,dark current \,less than \,0.001~e$^-$/ element/s at an operating temperature of 140~K~\cite{MacDougal2006}.

The practical application of CMOS sensors of this type was limited by significant (compared to CCD devices) element-by-element heterogeneity and problems with data calibration. For about a decade, they have been positioned as systems for auxiliary tasks that do not require high photometric accuracy~\cite{Simms2007, Bai2008, Bai2012, Downing2014}.

%=================================================================
\vspace{.7\baselineskip}
\noindent{\textbf{3.3 Back-illuminated CMOS sensors}}

\noindent{In 2014, E2V Technologies announced the creation of CIS113, a large-format three-way buttable CMOS sensor. Initially, the device had front illumination; later, a back-illuminated version was released. The device consists of 1920~$\times$~4608 16-$\mu$m elements, and has eight independent outputs that can sample individual areas of the device at a frequency of 20 Hz. A readout noise of no worse than 5~e$^-$ and a low dark current are provided. Due to back-illumination technology, the quantum detection efficiency reaches 90\%~\cite{Jorden2014}.}

Such characteristics, combining the capabilities of CMOS detectors and CCD technologies, were driven by the requirements of the TAOS-II project: a blind survey of eclipses to measure the size of trans-Neptunian objects in the range from 300~m to 30~km. TAOS-II will observe up to 10,000 stars with a frequency of 20 Hz using concurrently three telescopes with 1.3-m mirrors in a field with a diameter of $1\mbox{.\kern -0.7ex\raisebox{.9ex}{\scriptsize$\circ$}}7$~\cite{Lehner2012}.

Due to direct row and column addressing capabilities, 8~$\times$~8 element rectangular regions in each sensor can be sampled at a frequency of 20 Hz or higher at rates of up to 1 million elements/s per channel. This will enable conducting up to 20 billion photometric measurements per night and plotting approximately 70 million light curves per year.

To implement the project, for each telescope, a CMOS mosaic with cryogenic cooling was made from 10 CIS113 devices with a size of 9K~$\times$~9K elements~\cite{Wang_2022}.

%=================================================================
\vspace{.7\baselineskip}
\noindent{\textbf{3.4 Large-format scientific-grade CMOS detectors}}

\begin{figure}
\centering
\includegraphics[width=0.8\linewidth]{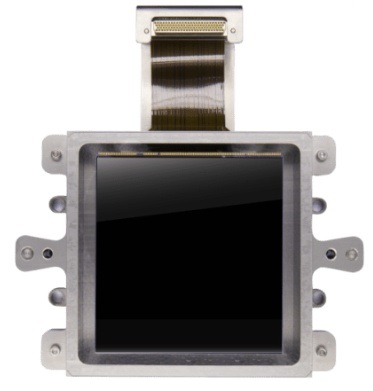}
\caption{\small{GSENSE 1081BSI CMOS device made of 9K~$\times$~9K elements in a silicon carbide frame.}}
\label{fig_08}
\end{figure}

\noindent{Founded only in 2012, Gpixel (PRC) has confidently won a share of the market of large-format detectors by releasing a line of CMOS devices under the GSENSE trade name by 2022. Of practical interest for astronomy are the devices GSENSE4040 with a 4K~$\times$~4K format of 9-$\mu$m elements, GSENSE6060 with a 6K~$\times$~6K format of 10-$\mu$m elements, and GSENSE1081BSI\footnote{\url{https://www.gpixel.com/products/area-scan-en/gsense/gsense1081bsi}} with a 9K~$\times$~9K format of also ten-micrometer elements (Fig.~\ref{fig_08}). These devices, except for the last one, are available in versions with front and back illumination, whereas GSENSE 1081BSI is only offered with back illumination. Because of the anti-reflection coatings, models with back illumination provide a quantum detection efficiency above 95\% at a sensitivity peak at a wavelength of approximately 600 nm.}

These devices have a number of features that make them preferable candidates for solving specific problems. For example, GSENSE4040 and GSENSE6060 have smaller linear dimensions, may be cooled to only $-55^\circ$C, and have noticeable dark currents of approximately 0.02~e$^-$/s per element, but can operate at frame frequencies of more than 20 Hz with a noise reading of approximately 2.5~e$^-$. However, GSENSE1081BSI, due to deeper cooling (to $-70^\circ$C), features five times less dark current and maintains three-side close butting, although it is just over half as good as younger models in terms of readout noise and a full frame is read in three seconds.

GSENSE line devices can supplant CCD detectors in most astronomical applications, even with the most demanding methods~\cite{Shen2022}. All these models can be cooled using Peltier elements, which simplifies their operation.

The main unsolved problem for CMOS devices is providing a dynamic range of the output signal comparable to that of CCD detectors. To realize the full dynamic range, each element is read twice with a different gain value, and then these readings are `glued together'. In this procedure, the accuracy of the result is usually deteriorated.

%=================================================================
\vspace{.7\baselineskip}
\noindent{\textbf{3.5 CCD and CMOS synthesis-LACera\texttrademark\\ from Teledyne Imaging}}

\noindent{In recent years, Teledyne Imaging has confirmed its claim to leadership in the creation of high-end optical detectors by announcing the development of LACera\texttrademark\ technology, which combines the advantages of CCD and CMOS detectors~\cite{Melle2021}.}

As noted above, as these technologies have been competing in recent decades. CMOS detectors have supplanted CCDs, especially in scientific applications involving low-light objects and the need for short exposures. Although CCD technology is still preferred for applications such as spectroscopy and astronomy, it is limited by relatively higher readout noise and low frame rates (with the exception of electron-multiplication CCDs). CMOS technology can overcome these limitations, but designing ever larger devices also encounters technological limitations \,and \,results \,in \,slower \,and \,noisier \,devices (such \,as GSENSE~devices; see above).

Teledyne Imaging successfully combined the benefits of CCD \,technology \,with \,those \,of \,CMOS \,sensor \,architecture. LACera\texttrademark\ technology provides deep cooling and low noise for large-format sensors, CCD-like back-illumination and global shutter capabilities, a high dynamic range, and glow reduction technology. This technology combines the presence of individual amplifiers for each element (as in CMOS) and an analog-to-digital converter for a separate detector row (as in CCDs).

This \,technology \,has \,been \,integrated \,into an astronomy-optimized COSMOS large-format CMOS camera. Combining all the benefits of LACera\texttrademark\ technology in one fully integrated device, COSMOS is preferred for astronomical applications requiring higher frame rates, high sensitivity, and low noise, while offering a wide range of sensor formats with high sampling rates --- from 3K~$\times$~3K (frequency of 50~Hz) up to 8K~$\times$~8K (with a frequency of~10 Hz) with a 10-$\mu$m element. The readout noise achieved at such rates is 0.7~e$^-$, which is comparable to the parameters for a CCD with electronic multiplication. With 18-bit digitization of the result, a dynamic range of 94 dB was achieved.

Because of all of the above, COSMOS cameras are an ideal solution for ground-based astronomical applications, such as high time resolution studies, monitoring of near-Earth space, and compensation of atmospheric turbulence using adaptive optics and speckle interferometry.

%=================================================================
\vspace{1\baselineskip}
\noindent{\large \textbf{4. Development of imaging systems\\at the Special Astrophysical Observatory\\of the Russian Academy of Sciences}}\vspace{.7\baselineskip}

\noindent{Research and development pertaining to the application of pano-ramic photodetectors in astronomical observations has been carried out at the Laboratory of Advanced Developments of the Special Astrophysical Observatory of the Russian Academy of Sciences for more than 40 years~\cite{Markelov2000, Murzin2016, Ardilanov2020, Ardilanov2021, Afanasieva2023}. The laboratory’s specialists develop and produce cryostatic imaging systems for large-format, low-noise, scientific-grade solid-state photodetectors used as part of the light-detecting equipment of BTA, the largest Russian telescope with a 6-m mirror, and optical telescopes operated in Russia and other countries.}

The main areas of research are:

--- techniques to achieve high sensitivity of the photodetector (minimizing thermal-generation current and readout noise), which is needed in observing extremely faint objects;

--- methods for obtaining maximum stability and linearity of the light to digital samples transfer characteristic of a video channel in varying external conditions;

--- principles for designing universal controllers that can operate in conjunction with photodetectors of various types, formats, and configurations, including mosaic ones.

An \,analysis \,of\, the \,requirements \,set \,by \,the \,most \,high-precision astronomical research methods shows that the photometric reproducibility of an ensemble of obtained images in studying faint objects over several hours or even nights should be at least 0.002\%, which corresponds to a level of approximately 0.05\% for a single exposure. In their development, the laboratory staff is making efforts to achieve such indicators of accuracy.

The development of five generations of universal controllers of the DINACON family (DSP-based INtelligent Array CONtroller) has made it possible to create photodetecting systems based on detectors of various types: CCD matrices and mosaics based on them, CCDs with internal charge multiplication, fast hybrid CCDs (pnCCDs), matrix multiplexed IR detectors, etc. In particular, systems with photodetectors featuring high sensitivity in the red and near-IR spectral regions have been created~\cite{Ardilanov2020, Afanasieva2023}.

The main photodetectors that are currently the most accessible in observational astronomy (including at the Special Astrophysical Observatory of the Russian Academy of Sciences) are wide-format scientific-grade CCDs produced by E2V Technologies with a format of 2K~$\times$~2K or more. They are usually devices with a thin substrate with back illumination and those based on depleted silicon.

%=================================================================
\vspace{.7\baselineskip}
\noindent{\textbf{4.1 CCD systems based on EEV CCD42-40 device}}

\noindent{By 1999, a new generation CCD system had been developed at the Special Astrophysical Observatory of the Russian Academy of Sciences~\cite{Markelov2000}, which enabled imaging with minimal readout noise, measured the charge relief created by the light flux with the utmost accuracy and stability required for high-precision photometry, and provided flexible control of wide-format low-noise CCD matrices of arbitrary format and topology and mosaic detectors.}

The DINACON-I CCD controller implements innovative techniques to reduce readout noise and increase charge measurement accuracy by implementing digital matched filtering and real-time correction of the video signal. Its novelty lies in the ability to measure the spectral density of noise to determine the parameters of a digital filter and optimize the mode of the output MOSFET, and the use of a flexible multiprocessor architecture. In comparison with its predecessors, the new controller provided a twofold reduction in readout noise, an increase in the dynamic range, an order of magnitude reduction in the nonlinearity of charge-to-digital sample conversion, and a tenfold improvement in gain stability. All these parameters characterize the transfer characteristics of the video channel\cite{Howell2006}.

In 2002 and 2003, two CCD systems with an EEV CCD42-40 photodetector of 2K~$\times$~2K 13.5-$\mu$m element format were put into operation at the telescopes of the Special Astrophysical Observatory of the Russian Academy of Sciences. One of the systems is used as part of the SCORPIO-I multimode focal reducer at the primary focus of the 6m telescope BTA~\cite{SCORPIO}, and the second operates as part of the attached equipment of the 1m Zeiss-1000 telescope~\cite{Z1000_Phot}. The following accuracy indicators were achieved in these systems: readout noise of 1.7~e$^-$ at a rate of 18 K element/s, 2.4~e$^-$ at a rate of 55 K element/s; instability of the built-in zero of $<1$~e$^-$; and video-channel gain instability of 0.02\%.

%=================================================================
\vspace{.7\baselineskip}
\noindent{\textbf{4.2 Imaging systems based on E2V CCD42-90}}

\noindent{Since 2007, the BTA telescope has been operating large-format, highly sensitive CCD systems based on E2V CCD42-90 devices with a format of 4.5K~$\times$~2K and 13.5-$\mu$m elements. A CCD system with a CCD42-90 BI detector, featuring enhanced sensitivity in the blue region of the spectrum, was incorporated in the Main Stellar Spectrograph (MSS) installed at the Nasmyth-2 focus~\cite{Panchuk2016}. A CCD system with a CCD42-90 DD detector, which has enhanced sensitivity in the red region of the spectrum, was included in the SCORPIO-2 universal spectrograph~\cite{SCORPIO2} installed at the primary focus of the telescope. The systems were controlled using modernized DINACON-II controllers.}

In these systems, the following accuracy indicators were achieved: readout noise of 1.85~e$^-$ at a readout rate of 25 K element/s; 2.8~e$^-$ at a rate of 100 K element/s; instability of the built-in zero of 0.25~e$^-$; instability of the video channel gain of 0.02\%; and thermal generation current of 0.003~e$^-$/s/element at $-$125$^\circ$C. The quantum sensitivity of the devices corresponded to the certified values: 89\% at a wavelength of 650 nm (at peak for CCD42-90 DD) and 94\% at a wavelength of 650 nm (CCD42-90 BI).

Measurements of the nonplanarity of CCD detectors using the three-dimensional laser triangulation method showed that the CCD42-90 DD device has deviations relative to the plane of the photosensitive surface of about 2 $\mu$m and a slope relative to the base fitting plane of 14 $\mu$m, and the surface of the CCD42-90 BI features a pronounced dome-shaped nonplanarity with a deviation of 7 $\mu$m and a slope of 20 $\mu$m.

\begin{figure}
\centering
\includegraphics[width=0.9\linewidth]{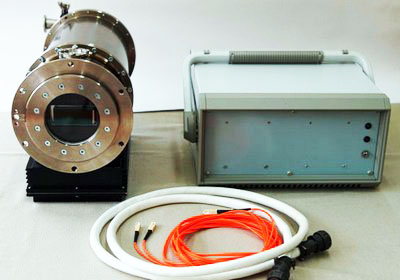}
\caption{\small{Imaging system based on the CCD42-90-1-G01 DD detector.}}
\label{fig_09}
\end{figure}

In 2018, a CCD system with a CCD42-90-1-G01 matrix with a thick substrate made of depleted silicon, which has a high sensitivity in the red region of the spectrum (Fig.~\ref{fig_09}), was \,installed \,on \,the \,NES \,high-resolution \,echelle \,spectrometer at the Nasmyth-2 focus. The system, which is operated by a DINACON-IV controller, has a Camera Link fiber-optic interface with a throughput of up to 6.8 Gbit s$^{-1}$. The photometric characteristics of the system are as follows: readout noise is 2.2~e$^-$ at a readout rate of 115 K element/s, maximum quantum efficiency of more than 90\% at a wavelength of 750 nm, thermal generation current at a temperature of $-$130$^\circ$C, 0.002~e$^-$/s/ element, and video channel nonlinearity is 1.0\%.

%=================================================================
\vspace{.4\baselineskip}
\noindent{\textbf{4.3 CCD systems for LAMOST telescope (China)}}

\begin{figure}
\centering
\includegraphics[width=0.9\linewidth]{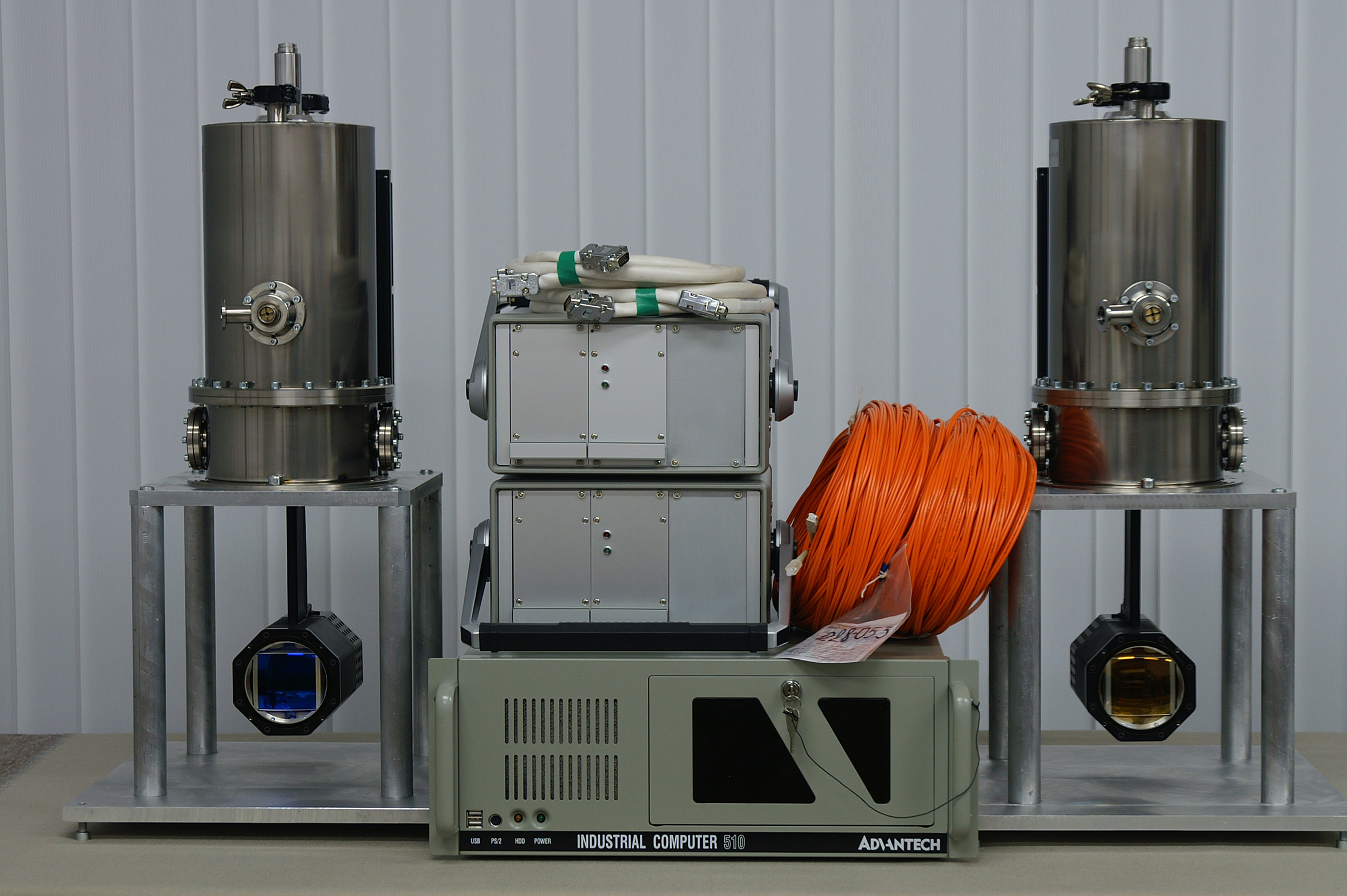}
\caption{\small{CCD systems for LAMOST telescope.}}
\label{fig_10}
\end{figure}

\noindent{In 2008, the Laboratory of Advanced Developments produced two pilot samples of CCD systems for the multi-object spectrograph LRS (Low Resolution Spectrograph) of the LAMOST telescope operated by the Chinese Academy of Sciences~\cite{Cui2012, Zou2006} (Fig.~\ref{fig_10}).}

The LRS is a dual spectrograph with dichroic beam division into two ranges: 370--590 nm (blue channel) and 570--900 nm (red channel). A feature of the optical design is the location of the photodetectors in the internal focus of the Schmidt camera. The design of the camera developed implements the required small profile of the camera’s separated optical head with a wide-format detector placed inside it and a corrective lens installed in front of it.

Two CCD203-82 CCDs manufactured by E2V Technologies were used as image detectors, with a format of 4K~$\times$~4K 12-$\mu$m elements with different coatings: one detector is sensitive in the blue region, while the second is sensitive in the red region of the spectrum (see Fig. 34 in~\cite{Cui2012}).

The photodetector is operated by a DINACON-III controller. Due to digital signal filtering in the video channel and video signal pre-amplifiers located in the camera head, low noise at a high readout rate is achieved. Various reading modes are implemented through four video outputs of the detector, and the following reading noise values are achieved: 3.3~e$^-$ at a rate of 50 K element/s and 8~e$^-$ at a rate of 1000 K element/s.

The nonplanarity of the wide-format detector is of extreme importance for the short-focus optical system of the LRS spectrograph. In this case, the range of nonplanarity of the detector surface should not exceed 30 $\mu$m. The actual nonplanarity of the `blue’ detector is 1 $\mu$m, and that of the `red’ detector is 8 $\mu$m.

%=================================================================
\vspace{.7\baselineskip}
\noindent{\textbf{4.4 Imaging system based on CCD231-84}}

\begin{figure}
\centering
\includegraphics[width=0.7\linewidth]{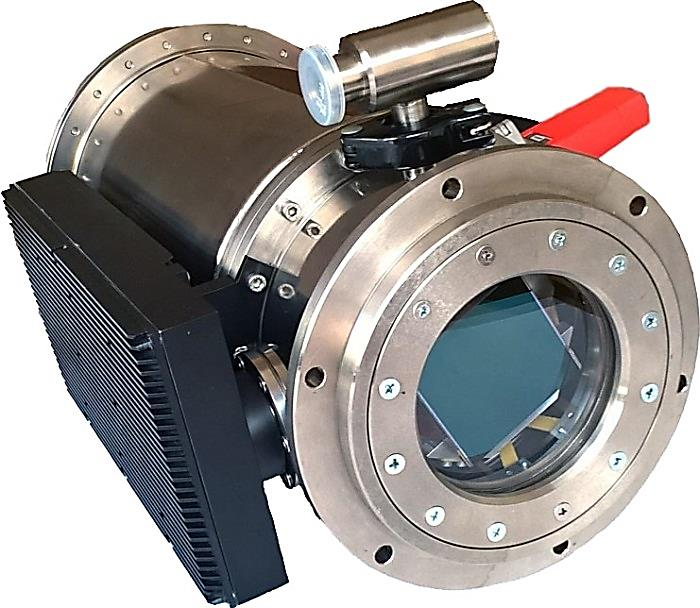}
\caption{\small{Cryostated camera with a CCD231-84 photodetector and camera electronics unit.}}
\label{fig_11}
\end{figure}

\noindent{Based on the fifth generation DINACON controller and an upgraded cryostat chamber with an enlarged entrance window, an imaging system was created in 2019 for a high-resolution fiber optic spectrograph installed on the BTA telescope (Fig.~\ref{fig_11}).}

This system uses a wide-format CCD matrix, CCD231-84, manufactured by E2V Technologies with a format of 4K~$\times$~4K back-illuminated 15-$\mu$m elements, a substrate made of high-resistivity silicon with a thickness of 40 $\mu$m, and an Astro Multi-2 coating. The photodetector has high sensitivity in the red and near-IR regions of the spectrum, with a peak quantum efficiency of 94\% at 750 nm. The readout noise is 2.48 e$^-$ at a rate of 135 K element/s, and the thermal generation current is 0.0005 e$^-$/s/element.

A distinctive feature of the photodetector is the large capacity of the potential well in the element (more than 200 Ke$^-$). which, along with low channel noise, provides a very high dynamic range of the signal in a single frame equal to 75\,000, or 97.5 dB.

Because the CCD camera with the CCD231-84 detector is stationary and located in a temperature-controlled room, it is possible to achieve high stability of the transfer characteristics of the controller’s video channels. Among the design features of the CCD camera, it can be noted that, for a light detector of this format, it was necessary to use an entrance window with a diameter of 120 mm made of 14-mm-thick coated quartz, and the distance from the inner surface of the window to the device was 6 mm.

The first scientific results obtained with the fiber-optic spectrograph of the BTA telescope using the created detection system confirmed both the high positional accuracy of Doppler measurements of the radial velocities of stars and the characteristics of the CCD camera declared by the developers~\cite{Valyavin2022}.

%=================================================================
\vspace{.7\baselineskip}
\noindent{\textbf{4.5 CCD system based on CCD261-84}}

\noindent{In 2021, a CCD system with enhanced sensitivity in the red region of the spectrum for a low- and medium-resolution multimode focal spectrograph began continuous operation at the BTA telescope~\cite{SCORPIO2}. This imaging system is the first to use a CCD261-84 detector with an extra-thick (200 $\mu$m) high-resi-stance silicon substrate. The main photoelectric characteristics, operating principles, and research results of the CCD system are described in~\cite{Afanasieva2023}.}

In 2022, the cryostatic chamber of this system was replaced with one with improved performance characteristics. The design of the chamber’s new nitrogen vessel reduces the risk of a cold leak, and the system for cleaning activated carbon from accumulated gasses enables quick resumption of cryopump operations without the need to depressurize the chamber. Due to the modernization of the thermal insulation system and the increased volume of the nitrogen vessel, the consumption of liquid nitrogen has been reduced, and the operating time of the cooling system from one refrigerant charge has been increased from 16 to 21 hours.

%=================================================================
\vspace{.7\baselineskip}
\noindent{\textbf{4.6 Photodetecting systems with GSENSE4040\\and GSENSE6060 detectors}}

\noindent{A study of the CMOS photodetector market shows that the previously mentioned GSENSE-series devices are more suitable for tasks requiring high time resolution than Teledyne E2V CMOS detectors (USA). Featuring comparable characteristics, scientific photodetectors of the GSENSE series are available for purchase, have a significantly lower price, and are simpler to control.}

As a result of the design and subsequent stages of development and manufacturing at the Special Astrophysical Observatory of the Russian Academy of Sciences, prototypes of the pCam4040 and pCam6060 photodetecting systems were created based on a CMOS controller and two large-format CMOS photodetectors: GSENSE4040CMT with front illumination and GSENSE6060BSI with back illumination. The architecture of the new controller, the structure of the two systems, and their characteristics are described in detail in~\cite{Ardilanov2021, Ardilanov2021_2}. Special software has been implemented~\cite{Afanasieva2015}, which makes it possible to control the cooling of the camera and all operating modes of the photodetector, read and save video data, perform telemetry of the operating parameters of the photodetector, and automate the observation process.

The photodetectors are cooled by a thermoelectric module. The connection to the control computer is maintained via a fiber-optic communication line at a distance of up to 300 m. The electronics of the controller and power supply are sealed according to the IP66 standard, which meets the requirements of astronomical observations carried out outdoors.

For a photometric system based on GSENSE4040 at an operating temperature of $-$25$^\circ$C in high gain mode, a readout noise of 4.4~$e^{-}$ and a dark current of 0.05~$e^{-}$/s/element were achieved. The nonlinearity of the transfer function was 1.2\%, and the dynamic range of the system in this mode was approximately 750. Because of the use of microlenses, the quantum detection efficiency for this device with frontal illumination was 74\% at a wavelength of 600 nm.

The conducted studies with the photodetector revealed a certain feature in the behavior of the photodetector in observing bright objects in modes with long exposure with the accumulation of a large amount of charge. Subsequent exposure frames contain a residual image of the previous bright object (see Fig. 2 in~\cite{Ardilanov2021}). This feature, which manifests itself in detectors with frontal illumination, is not typical of BI detectors.

The main advantages of the developed CMOS system are the high frame readout rate (two orders of magnitude higher than in CCD systems), high sensitivity (comparable to that of CCD’s), and a large detector format, which allows the use of this imaging system for a panoramic survey of celestial spheres, obtaining information about rapidly varying astronomical objects by spectral and photometric means for the speckle interferometry method, and compensating for wavefront distortions.

Currently, offered on the global market of photodetecting systems is a wide range of systems based on GSENSE \,detectors: \,Kepler \,KL4040, \,Kepler KL6060 \,(FLI, \,USA), QHY4040Pro, QHY6060 (QHY, PRC), PX4040A/B, PX6060 (PixelX, PRC), MX377 (XIMEA GmbH, Germany), NEVA4040, NEVA6060 (NPK Fotonika, Russian Federation), etc. The competitive advantages of the pCam4040 and pCam6060 cameras over their counterparts are the extended dynamic range (84 dB) and optimized combined image in high dynamic range (HDR) mode, the sealed camera design with IP66 protection degree, the reduced dark current mode during accumulation, and the availability of a 10 gigabit Ethernet communication interface with an optical cable length of up to 300 m.

%=================================================================
\vspace{.9\baselineskip}
\noindent{\large \textbf{5. Prospects for the development\\of large-format imaging systems\\for optical astronomy in Russia}}\vspace{.7\baselineskip}

\noindent{It is fairly apparent that the effectiveness of astronomical research largely depends on the capabilities of the imaging systems used. As shown above, all breakthrough astrophysical discoveries, both ground- and space-based, have employed all the potential capabilities of new radiation detectors in terms of accuracy, reproducibility of results, and information capacity. This applies to the geometric dimensions of detectors and mosaics constructed on their basis and to the obtainment of data with the highest photometric accuracy possible, time scale reference, etc.}

Therefore, researchers’ interest in systems created at the forefront of modern technologies is quite understandable. Until recently, almost the entire market for such products, with the exception of scientific-level infrared systems, was available to Russian researchers. Recent events and a certain monopolization of the market for scientific detectors for optical astronomy are closing these opportunities, hopefully temporarily. The reason is not pricing policy alone.

Developers and users of systems for the detection of optical radiation are now encountering the problem of choosing promising technical solutions. Most optical telescopes operated in Russia are equipped with systems based on CCD detectors of various classes. To upgrade them and equip the new instruments being created, new, and most often, large-format photodetecting systems should be implemented. It should be admitted that, since both large-format detectors and systems based on them produced by Western manufacturers are now unavailable, it is necessary to make a `turn to the East’ and reorient users to products manufactured either in Russia or in the PRC.

The requirements for photodetecting systems in modern experiments are currently met by only large-format CMOS devices of the GSENSE series.

Photodetecting systems based on these devices can be purchased from both Chinese manufacturers (QHY, PixelX, etc.) and Russian companies (NPK Photonics, Raster Technology, NPK SPP, SAO RAS). No information about CCD devices with similar parameters from Chinese manufacturers is available. Russia still has competence in the production of large-format CCD sensors: NPP ELAR has been producing such detectors since the late 1990s. However, their characteristics are still inferior to the world level, if only because of the lack of technologies for creating devices with back illumination and using thick substrates with depleted silicon. As a result, the efficiency of detecting weak optical flows is almost halved. This level was reached in the West almost 40 years ago, but it is extremely important for us to go through this stage in the coming years. Attention of the Russian Government should be drawn to this problem, which will solve the issue of enhancing the efficiency of scientific research both in Russia and in our BRICS partner countries, which also need to rely on their own developments.

%=================================================================
\vspace{1\baselineskip}
\noindent{\large \textbf{6. Conclusions}}\vspace{.7\baselineskip}

\noindent{Reviewed above are the implementation of state-of-the-art technologies for creating highly efficient large-format photodetecting systems of various types and identification of the main trends in global instrument making. They are illustrated using the example of the most efficient scientific instruments developed for use in optical astronomy. Large-format systems created by the authors over the past 20 years at the Special Astrophysical Observatory of the Russian Academy of Sciences for telescopes operated in Russia and abroad have been briefly analyzed, and it has been shown that, in terms of their operational parameters, they are not inferior to their foreign counterparts. The problem of creating new promising imaging systems in Russia using the nation’s own capabilities is discussed.}

Part of the observational data was collected using the unique scientific facility Big Telescope Alt-azimuthal of SAO RAS, and data processing was performed with the financial support of grant No 075-15-2022-262 (13.MNPMU.21.0003) of the Ministry of Science and Higher Education of the Russian Federation.

The figures~\ref{fig_01}--\ref{fig_08} were taken from open sources in the Internet network, the figures ~\ref{fig_09}--\ref{fig_11} were presented by the authors.

%\selectlanguage{russian}

%=================================================================

\end{document}